\documentstyle[epsf]{article}

\begin{document}

\pagestyle{empty}


\title{On velocities beyond the speed of light $c$}

\author{ Giani Simone \\ CERN, Geneva - CH }

\maketitle

\begin{abstract}

From a mathematical point of view velocities can be larger than $c$.
It has been shown that Lorentz transformations are easily extended in Minkowski
space to address velocities beyond the speed of light.
Energy and momentum conservation fixes the relation between masses and
velocities larger than $c$, leading to the possible observation of negative mass 
squared particles from a standard reference frame.
Current data on neutrinos' mass square yeld negative values, making neutrinos
as possible candidates for having speed larger than $c$.
In this paper, an original analysis of the SN1987A supernova data is proposed.
It is shown that all the data measured in '87 by all the experiments are 
consistent with the quantistic description of neutrinos as combination of 
superluminal mass eigenstates.
The well known enigma on the arrival times of the neutrino bursts detected at 
LSD, several hours earlier than at IMB, K2 and Baksan, is explained naturally.
{\em It is concluded that experimental evidence for superluminal neutrinos was
recorded since the SN1987A explosion, and that data are quantitatively 
consistent with the introduction of tachyons in Einstein's equation}. 

\end{abstract}


\section{Theory}

In Minkowski's space, Lorentz transformations are expressed as rotations
in the plane $x_{1}x_{4}$, where $x_{1}$ is the space coordinate associated to
the direction of the relative velocity $v$ between two reference systems, 
and $x_{4}$ is equal to $ict$ \cite{Minkowski}.
A relative velocity of $c$ corresponds to
a rotation dividing the first quadrant of the complex plane $x_{1}x_{4}$ into 
two equivalent parts.
However, both velocities greater than $c$ (tachyons) \cite{Feinberg} and lower 
than $c$ correspond to 
valid rotations (of course ending on different sides of the diagonal defined by 
$v=c$). The Lorentz transformations
\begin{equation}
x_{1}' = \frac{x_{1} + i x_{4} v / c}{\sqrt[2]{1 - v^{2} / c^{2}}} \; , \;\;\;
x_{4}' = \frac{x_{4} - i x_{1} v / c}{\sqrt[2]{1 - v^{2} / c^{2}}}
\end{equation}
show that the effect of setting $v > c$ is to swap the roles (real vs imaginary)
of $x_{1}'$ and $x_{4}'$ \cite{Mignani}. 
In this sense, the speed of light is a threshold 
velocity, but not the maximum allowed velocity \cite{Olkhovsky}. 
The speed of light remains the
only velocity which is constant in any reference system, making the invariant
$s^{2} = \Delta x_{1}^{2} + \Delta x_{4}^{2}$ equal to 0. Particles with
velocities lower/greater than $c$ have a negative/positive $s^{2}$
(and different velocity for different reference systems).
From this discussion, it does not follow that it is possible to accelerate a
particle in vacuum from a velocity lower than $c$ to velocities greater than 
$c$ (similarly, it is not possible to decelerate a photon in vaccum to a
speed lower than $c$). However it is argued that special Relativity does not
exclude \cite{Bilaniuk}, \cite{RecamiMignani} the possibility for particles 
to exist in three different regimes of velocities $V$: 
\newline $\; 0 \leq V < c$ ; $\; V = c$ ; $\; c < V < \infty$.

All known particles and interactions are observed to conserve energy and
momentum. If any known particle has speed larger than $c$, this should
not affect the (measured) energy and momentum balance in the interactions 
that are undertaken by such a particle. The particle energy is given by 
\cite{Einstein}
\begin{equation} 
E = \frac{m_{0} c^{2}}{\sqrt[2]{1 - v^{2} / c^{2}}}. 
\end{equation} 
If the velocity $v$ is greater than $c$, the denominator would become imaginary,
but if, at the same time, the mass (intended here and in the following as rest
mass observable from an Earthly reference frame) can be observed as imaginary, 
the value of the energy is real \cite{Frewin}, \cite{Hook}.                                          
Particles with velocity larger than $c$ are observable as with imaginary rest 
mass and, viceversa, particles observable as with imaginary rest mass have 
velocity larger than $c$ \cite{Chiao}.
A complete treatment of generalized Lorentz transformations \cite{Recami2D} shows how
tachyons can have real rest mass and their own rest reference frame, via a 
change of sign in the quadratic forms of the relativistic invariants occurring when 
transforming from subluminal to superluminal reference frames or viceversa.

In all cases, equation (2) becomes equivalent to:
\begin{equation}
E = Re (\frac{m_{0} c^{2}}{\sqrt[2]{1 - v^{2} / c^{2}}}) \; ; \;\;\; 
0 = Im (\frac{m_{0} c^{2}}{\sqrt[2]{1 - v^{2} / c^{2}}}).
\end{equation}

Consequences \cite{Recami}:
\begin{itemize}
\item $m_{0}^{2}$ could be observed to be negative.
\item The greater (assuming it is always larger than $c$) is the velocity, 
      the lower would be the particle's energy. It would remain true that 
      the energy tends to infinity when the velocity tends to $c$. 
\item For a given energy, heavier masses imply higher superluminal velocities.
\end{itemize}

\section{Experimental data}

Existing experimental data on neutrinos seem to be consistent with the 
hypothesis that their mass square is negative \cite{Chodos}
(see also \cite{Chodos}, \cite{Kostelecky}, \cite{Rembielinski}, \cite{Mestres} 
for several theoretical treatments of superluminal neutrinos). 
Seven independent measurements of the electron neutrino mass square are 
reported in \cite{Peltoniemi}. Six of them yeld a negative value for the 
result, 
though each experiment reports of large statistical and systematic errors. 
Their range is from $-147 \pm 68 \pm 41 (eV / c^{2})^{2}$ to 
$1.5 \pm 5.9 \pm 3.6 (eV / c^{2})^{2}$.
In addition, it has been computed \cite{Langacker} that the weighted average 
of the 
electron neutrino mass square measured in tritium $\beta$ decay experiments is
$m_{\nu _{e}}^{2} = -96 \pm 21 (eV / c^{2})^{2}$.
Finally, the best fit of the measurements considered by the Particle Data
Group \cite{Pdg} gives $m_{\nu _{e}}^{2} = -27 \pm 20 (eV / c^{2})^{2}$ 
(see \cite{Pdg} also for a preliminary discussion on the relation between the 
electron
neutrino mass and the values associated to the neutrino mass eigenstates).
Concerning the muon neutrino mass square, two independent measurements find
a negative value \cite{Pdg}, though given the reported errors it is safer to 
give only an upper limit for the mass \cite{Pdg}.

\section{SN1987A data: the experimental signature of superluminal neutrinos?}

At the level of astrophysical data, a possible signature of the superluminal 
nature of neutrinos can be found. 
An original analysis of the neutrino bursts
detected in 1987, before the optical observation of the SN1987A supernova,
is proposed here. A summary of the relevant data follows below:
\begin{itemize}
\item The distance between the Earth and SN1987A (in the Large Magellanic 
      Cloud) is computed to be 170000 light-years \cite{Audouze}, \cite{Hughes},
      \cite{Rossman}, \cite{Soper} (though
      the values of 160000 and 180000 light-years are found as well 
      \cite{Cramer}).
\item Sophisticated and different models of stars implosion and supernovae
      explosion agree on the fact that neutrinos are emitted within an interval
      of a "few" seconds (peaked, especially for electron neutrinos, at a few
      hundreds of milli-seconds, and consistently also with the eventual 
      neutrino heating mechanism) \cite{Burrows}, \cite{Mayle}, 
      \cite{Mezzacappa}.
\item Indeed, several detectors revealed neutrino bursts, consistent with 
      electron neutrinos events \cite{Hughes}, \cite{Petschek}, on 23rd 
      February 1987:
      \begin{itemize}
      \item IMB found 8 events within 5.58 sec, with energy between 19 MeV and 
            39 MeV \cite{Audouze}, \cite{Hughes}, \cite{Burrows}, \cite{DeRujula} 
            (errors between 5 MeV and 9 MeV).
      \item K2 found 12 events (1 discarded later) within 12.439 sec, with
            energy between 6.3 MeV and 35.4 MeV \cite{Audouze}, \cite{Hughes},
            \cite{Burrows}, \cite{DeRujula} (errors between 1.7 Mev and 8 MeV).
      \item Baksan found 5-6 events \cite{Audouze}, \cite{Hughes} with energy 
            between 12 MeV and 23.3 MeV.
      \end{itemize}
      The syncronization between the three experiments was not better than
      about one minute \cite{Audouze}, \cite{Hughes}, though all neutrinos are 
      normally considered
      to have arrived within about 15 sec. It has to be noted also that the
      energy values reported above refer to the electrons which produced the
      Cherenkov light signal in the detectors: whenever the detected Cherenkov
      light was instead produced by a positron, the neutrino energy should 
      be increased by about 2 MeV \cite{Petschek}.
\item Supernova models predict that the emission of visible light follows the
      neutrino burst after a few hours (for blue giants, like in the case
      of the SN1987A progenitor) \cite{Soper}, \cite{Petschek},
      or after ten(s) hours (for red giants) \cite{DeRujula}, \cite{Petschek}: 
      in fact the explosion 
      of the supernova occurs when the shock-wave generated by the material 
      rebounded by the neutronized core, reaches the outer star layers.
\item Indeed, it has been possible to reconstruct that the first photografic
      observation of SN1987A was recorded between 09h:36min and 10h:38min GMT,
      while the neutrino burst was observed at 07h:36min GMT \cite{Audouze},
      \cite{Soper}, \cite{Mezzacappa}.
      Hence the delay was within 2 and 3 hours.
\end{itemize}
The picture seems consistent, however there is an experimental observation 
which is currently un-explained: the LSD experiment at Mont Blanc observed 
5 events within 7 sec, with energy between 5.8 and 7.8 MeV 
\cite{Hughes}, \cite{DeRujula} (errors of about 1-2 MeV
according to \cite{DeRujula}, energy resolution of about 15\% or 25\% according
to \cite{Saavedra}), 
earlier than K2 and IMB by 4h:43min:4.58sec \cite{Hughes}, \cite{DeRujula}.
The events were consistent with electron neutrinos events \cite{Hughes}, 
\cite{Petschek}.
A thourough analysis showing that those events could not be due to
statistical noise was performed in \cite{DeRujula}, also combined with the 
fact that K2
recorded at least one event 4h43min13$ \pm 2$sec before the published 
burst \cite{DeRujula}, \cite{Schramm}.
It should also be noted that IMB could not have recorded events at those 
energies because of its higher thereshold. Moreover, it should be considered
that LSD detected 2 candidate events at IMB+K2 time, of energies between 7 and
9 MeV within an interval of 13 sec \cite{Saavedra}, \cite{Schramm}.

In order to analyse the data, 
it is recalled that the so-called electron-neutrino, muon-neutrino
and tau-neutrino are eigenstates of the weak interaction, each of them being
a linear combination of mass eigenstates \cite{Kayser}, \cite{Giunti},
\cite{Bilenky} (assuming neutrino masses different from zero):
\begin{equation}
\nu_{f} = \sum_{m} U_{fm} \nu_{m},
\end{equation}
where $f$ refers to the flavour and $m$ to the mass eigenstates.
A proper wave-packet treatment (including spreading) of the neutrino 
\cite{Kayser}, \cite{Giunti},
\cite{Mohanty} emphasises that each mass eigenstate has its own energy and 
momentum \cite{Mohanty} 
(the momentum difference between eigenstates cannot be neglected for extremely 
long distances or times-of-flight, i.e. for supernova neutrinos). 
This leads to different speeds for the mass eigenstates and hence to their 
eventual separation after long enough distances \cite{Kayser}, which could be 
observed for neutrinos coming from supernovae \cite{LoSecco}.
In any case, in regime of non-oscillations as well, any flavour can be selected 
(with a given probability) even during the detection of a single mass eigenstate
\cite{Kayser}, \cite{Giunti}, \cite{Mohanty}, as it can be easily 
guessed inverting equation (4). Hence this explains the reason why all the
detected events can be consistent with neutrinos of the electron flavour.
Then the following approach is proposed to explain the enigma of the LSD events:
\begin{itemize}
{\em
\item All the SN1987A neutrinos have been emitted in a "few" seconds.
\item The faster superluminal mass eigenstate of all neutrinos reached the 
      Earth about 4h43min before the corresponding slower superluminal mass 
      eigenstate.
\item LSD (and K2 with one event) detected the faster superluminal mass 
      eigenstate (at the corresponding energies) of the neutrinos, while IMB, 
      K2 and Baksan (and LSD with two events) detected their slower superluminal
      mass eigenstate (at the corresponding energies).
}
\end{itemize}
The mass eigenstate $|m1\!\!>$ is assumed to be closely associated to the
electron-neutrino flavour (so the measured $m_{\nu_{e}}$ will be used for it), 
while the mass eigenstate $|m2\!\!>$ is assumed to be closely associated to the 
muon-neutrino flavour (so $m_{\nu_{\mu}}$ will be used for it) \cite{Pdg}. 
The tau-neutrino is for the moment neglected. However, it cannot be 
excluded that LSD detected $|m3\!\!>$, more closely associated to the 
tau-neutrino, rather than $|m2\!\!>$. 
Therefore, in the following, the notations $\mu$ and $|m2\!\!>$
should be considered as indicating either $\mu$ or $\tau$ and $|m2\!\!>$ or
$|m3\!\!>$.

For a quantitative treatment of the approach outlined above, the first step is 
to show that the IMB+K2 burst can correspond to the detection of the $|m1\!\!>$
mass eigenstate of the SN1987A neutrinos.
Equation (3) is applied, at first, to the mass eigenstate $|m1\!\!>$, setting 
the mass to be $m_{\nu_{e}}^{2} = -27 (eV/c^{2})^{2}$, and varying the 
energy between
the limits of the measured spectrum, from $E_{min}$ = 6.3 MeV (K2) to $E_{max}$ 
= 39 MeV (IMB).
Then, the corresponding velocities can be computed. These are used to compute
the time delays introduced by the different energies within the burst of
neutrinos of mass eigenstate $|m1\!\!>$ (using the distance from SN1987A to the 
Earth to compute the time differences $\Delta T$).
The results (expressed in terms of the time gained vs hypothetical neutrinos 
travelling at the speed of light) are:
\begin{equation}
\Delta T(E_{min}^{\nu_{e}} ; Light) = 1.8 sec \; ; \; 
\Delta T(E_{max}^{\nu_{e}} ; Light) = 0.05 sec.
\end{equation}
This gives two interesting consequences:
\begin{itemize}
\item For the $|m1\!\!>$ neutrinos eigenstate, the computed spread of the 
      arrival times due to different velocities (energies), found to be lower 
      than 2 sec, is well contained in the observed time intervals at IMB and 
      K2. This means that basically it does not affect the arrival time spread 
      due to the emission spread of the SN1987A neutrinos, and hence is 
      consistent with the observations (the spread due to different velocities
      which has been computed above can indeed allow to improve the
      fits of the K2 data). Anyway note that, for the moment, similar
      results could be obtained by applying equation (2) with a real mass and
      velocities lower than $c$.
\item However, if it is true that $|m1\!\!>$ is the eigenstate seen at 
      IMB+K2 time,
      the $|m2\!\!>$ eigenstate, in order to have arrived 4h:43 min earlier,
      must definitively have traveled at a speed greater than $c$. 
      This is also consistent with an heavier mass for $|m2\!\!>$
      (as from equation (3)).
\end{itemize}
Therefore, the second step in this quantitative treatment is to show that the
LSD burst is consistent with the detection of the $|m2\!\!>$ eigenstate of the
SN1987A neutrinos.
If equation (3) is applied using an average energy measured at LSD
(= 6.8 MeV) and using the velocity needed to justify an advance of 4h43min
over the $|m1\!\!>$ eigenstate, it is possible to compute a value for the mass
associated to the $|m2\!\!>$ eigenstate. The result is of the order of
\begin{equation}
m_{\nu_{\mu}}^{2} \simeq -(541 eV/c^{2})^{2},
\end{equation}
which is not exluded, for suitable mixing angles, by most of the neutrino 
oscillation experiments \cite{Mohanty}, \cite{Grimus}, nor by astrophysical
considerations \cite{Giannetto}.
The reason why an average energy measured at LSD can be used is shown below: 
\begin{itemize}
\item The mass and velocities needed to satisfy equation (3) and, at the same
      time, the requirement of $\Delta T$ = 4h:43min (before $|m1\!\!>$ arrival)
      are such that the energy of $|m2\!\!>$ is required to be constant within 
      about 0.1\%, if the time spread due to different velocities (energies) 
      has to be contained in the observed time spread (7 sec) between LSD 
      events.
\item Indeed, differently from the IMB and K2 events, the 5 LSD events can be
      easily addressed by a fit at constant energy, as it can be evidently 
      deduced from the LSD data and relative errors reported above. 
      In addition, if the K2 event measured at LSD time is considered, its 
      energy of 7 MeV \cite{Schramm} is also consistent with the fit at 
      constant energy.
\end{itemize} 
A proper error analysis goes beyond the scope of this paper, since the
result on the mass eigenvalue is sensitive to the uncertainties on several 
quantities, such as on the errors on the energy of the LSD (K2) events and on 
the exact distance from SN1987A; hence an average of the LSD energies is used 
for a proof-of-concept, rather than trying to perform a proper fit (or even to
compute the arithmetic mean). 
In addition, it should be observed that LSD has also published \cite{Saavedra} 
a burst of 4 events (within a 57 sec interval), detected about 42min even 
earlier than the previously analysed burst. According to equation (3), if they
correspond to the detection of the same mass eigenstates as before, their 
energy should not be different (smaller) from 6.8 MeV by more than about 6\%, 
and indeed those data points can fit the requirement, if properly considering 
their error reported in \cite{Saavedra}. 
It is also interesting to consider that events at much lower energies can have
arrived to Earth much earlier than the measured bursts, but they cannot
have been detected because of the energy threshold of the experiments.
Finally, note that no attempt is made here to investigate the relations
between the observed energy of the mass eigenstates and the energies
at which neutrinos were produced in the supernova (which imply model-dependent
considerations and red-shifting plus eventual gravitational effects), 
since only time-intervals and energies meaured on Earth are consistently used 
in the computations. 

So far, it has been demonstrated that experimental data are consistent with the
hypothesis that LSD (and probably K2) detected the faster and 
heavier $|m2\!\!>$ 
mass eigenstate, while IMB, K2 and Baksan detected the slower and lighter
$|m1\!\!>$ mass eigenstate of a single superluminal neutrino burst emitted by
SN1987A in a few seconds. 

However, it has not been excluded yet the possibility
that LSD detected a faster and lighter $|m1\!\!>$, while IMB, K2 and Baksan
detected a slower and heavier $|m2\!\!>$, both associated to real masses, of a 
single burst of neutrinos traveling at velocities lower than $c$, according to
equation (2). Anyway, this alternative option is discarded because of the
following arguments:
\begin{itemize}
\item Applying equation (2) to the supposed slower and heavier (real mass)
      $|m2\!\!>$, the spread of energies measured at IMB+K2 (and the consequent
      spread of velocities) provokes a spread of arrival times much bigger
      ($> 20 $ hours) than the observed one ($\approx 15 $ sec, certainly 
      $< 1 $ minute).
\item Since the speed difference between the $|m1\!\!>$ eigenstate 
      (at LSD energy)
      and $c$ is negligible, and since the optical recording of the 1987A
      explosion occurred 2-3 hours after IMB+K2 time, it follows that the
      first photons emission of SN1987A would have started about 7-8 hours
      after the neutrinos emission. This is not favoured (compared with a 2-3
      hours delay) by the current modelling of a blue giant such as the 
      progenitor of 1987A. 
\item Though the statistics is low, all the neutrino events detected from
      1987A are consistent with electron-neutrino flavour \cite{Hughes},
      \cite{Petschek}. Hence it
      is more likely that the $|m1\!\!>$ eigenstate was revealed at IMB+K2 time
      because, overall, more neutrinos were detected there than at LSD time.
      However, it would be necessary to take into account the mixing angles, 
      the detectors' thresholds and sensitivities, and to have sufficiently 
      high statistics, in order to be able to use this argument.
\end{itemize}
Final considerations can be done about the reasons why LSD detected more
events than K2 at LSD time: statistical fluctuations can explain this, but
also the fact that the energies were around 6 MeV, where LSD was more efficient
than K2 (because of the lower threshold), can be an explanation. Obviously,
at IMB+K2 time, the energies involved were higher and the larger fiducial
volumes of K2 justifies its higher counts.

\section{Future Research}

From a theoretical perspective, it would be interesting to re-analyse the
concept of simultaneous events in the framework of Special Relativity when
considering velocities greater than $c$ \cite{RecamiWaldyr}, \cite{Eddington}.  
In addition, some gravitational effects,
in particular for black-holes, could be revisited in view of the existence of
superluminal neutrinos of known mass.

From an experimental perspective, it would be interesting to perform direct
measurements of the neutrino velocities in vacuum, to be compared with $c$.
According to equations (3) and (6), a $|m2\!\!>$ neutrino eigenstate of 6.8 MeV 
energy should have a velocity approximatively equal to 
3.0000000095047575e+05 km/sec.
For distances of the order of the Earth-Sun distance, this gives time
differences (between neutrinos and light) of the order of micro-seconds. 
An Earth-based accelerator and a spacecraft-based detector (or viceversa)
seem the only reasonable fit for such an experiment.
However, managing to generate and detect neutrinos of energy equal to 68 KeV,
would allow the observation of time differences (between neutrinos and light)
of the order of a micro-second for distances of the order of the Earth radius. 
This would make it possible to use existing accelerators and Earthly detectors
(or detectors on satellites) to measure the neutrinos velocities. 
Another interesting experience could be related to the deflection angles of
neutrinos by strong gravitational fields. 

In conclusion, if some estimation of $m_{\nu_{\tau}}$ would be available (and in
case it is assumed that LSD detected the mass eigenstate closer to $\nu_{\mu}$),
it would be interestng to scan backwards the IMB, K2 and 
LSD data, looking for signals at times prior to the LSD time.
Viceversa, if the LSD burst corresponds to the detection of the mass eigenstate
closer to $m_{\nu_{\tau}}$, then it would be interesting to scan the
time interval between LSD and IMB+K2 times, searching for a burst that might
correspond to the mass eigenstate closer to $\nu_{\mu}$.
Alternatively, if two of the mass eigenstates would have very close masses,
they could have appeared in the same burst (at LSD or IMB+K2 time depending
on which eigenstates are considered).
Finally, individual events during the LSD -- IMB+K2 interval, could account
for neutrinos of the same mass eigenstate as detected at LSD, but with higher
energies.

\section{Conclusions}

Particles with velocities greater than $c$ and negative observable mass square are 
allowed by the conservation of the relativistic energy-momentum and generalized 
Lorentz transformations.
Current data for neutrinos are consistent with this hypothesis.

{\em The expression of neutrinos as combination of superluminal mass eigenstates
is consistent with all data (including LSD) recorded from the SN1987A
supernova.
Their analysis shows that the SN1987A data can thus be the experimental
verification of the tachyons theory.}

Direct measurement of the neutrino velocities in vacuum at different energies 
could further confirm the theory or definitively reject the neutrino as a 
candidate for superluminal velocities.


\begin{thebibliography}{99}

\bibitem{Minkowski} H.Minkowski, "Space and Time", 
         80th Assembly of German Natural Scientists and Physicians, 1908.

\bibitem{Feinberg} G.Feinberg, "Particles that go faster than light",
         Scientific American, 1970.

\bibitem{Mignani} R.Mignani et al., "Superluminal Frames and Group of 
                                     Generalized Lorentz Transformations in Four
                                     Dimensions",
         Lettere al Nuovo Cimento, 1972.

\bibitem{Olkhovsky} V.S.Olkhovsky et al., "About Lorentz Transformations and 
                                           Tachyons",
         Lettere al Nuovo Cimento, 1971.

\bibitem{Bilaniuk} O.M.P.Bilaniuk, E.C.G.Sudarshan, Am. Journal of Physics
         1962, Physics Today, 1969.

\bibitem{RecamiMignani} E.Recami et al., "Superluminal Inertial Frames in 
                                          Special Relativity",
         Lettere al Nuovo Cimento, 1973.

\bibitem{Einstein} A.Einstein, "Vier Vorlesungen uber Relativitatstheorie",
         Friedrich Vieweg \& Sohn, Braunschweig, 1922.

\bibitem{Frewin} R.Frewin et al., "Superluminal motion",
         Ketav, 1997.

\bibitem{Hook} W.H.Hook, "What are tachyons? Are they real?",
         [Sci.Astro] Astrophysics, 1997.

\bibitem{Chiao} R.Y.Chiao et al., "Faster than light?",
         Scientific American, 1993.

\bibitem{Recami2D} E.Recami, "Classical Tachyons and Possible Applications",
         Rivista del Nuovo Cimento, 1986.

\bibitem{Recami} E.Recami et al., "Tachyons, monopoles and related physics"
         North Holland, 1978.

\bibitem{Chodos} A.Chodos et al., "The neutrino as a tachyon",
         Phys.Lett.B 1985.
        
\bibitem{Kostelecky} V.A.Kostelecky, "Mass bounds for space-like neutrinos",
         World Scientific Singapore, 1993.
         
\bibitem{Rembielinski} J.Rembielinski, "Tachyonic neutrinos?",
         hep-th/9411230, 1994.
         
\bibitem{Mestres} L.G.Mestres, "Space, time and superluminal particles",
         Physics 9702026, 1997.

\bibitem{Peltoniemi} J.Peltoniemi, "Laboratory measurements and limits for 
                                    neutrino properties", 
         1997.

\bibitem{Langacker} P.Langacker, "Implications of neutrino mass",
         University of Pennsylvania, 1995.

\bibitem{Pdg} Physical Review D, "Particles and Fields",
         The American Physical Society, 1996.

\bibitem{Audouze} J.Audouze et al., "The Cambridge Atlas of Astronomy"
         Cambridge, 1994.

\bibitem{Hughes} I.S.Hughes, "Elementary Particles",
         Cambridge, 1991.

\bibitem{Rossman} J.Rossman, "The SuperNova SN1987A - Stellar Dynamics",
         P.U.R.I., 1997.

\bibitem{Soper} D.E.Soper, "The story of Supernova 1987A",
         University of Oregon, 1997.

\bibitem{Cramer} J.G.Cramer, "SN1987A - Supernova Astrophysics",
         AVC AV-23, 1996.

\bibitem{Burrows} A.S.Burrows, "Neutrinos from Supernovae",
         A.G.Petschek Springer-Verlag, 1990.

\bibitem{Mayle} R.W.Mayle, "Neutrino Heating Supernovae",
         A.G.Petschek Springer-Verlag, 1990.

\bibitem{Mezzacappa} A.Mezzacappa et al., "A computational Grand-Challenge
                                           Modelling Core-Collapse Supernovae",
         Proposal DOE, 1996.

\bibitem{DeRujula} A.DeRujula, "May a supernova bang twice?",
         Physics Letters B, 1987.

\bibitem{Petschek} A.G.Petschek, "Supernovae",
         Springer-Verlag, 1990.

\bibitem{Saavedra} O.Saavedra, "Neutrino astronomy at Mont Blanc: from LSD
                                   to LSD-2",
         LaThuile proceedings, 1988.

\bibitem{Schramm} D.N.Schramm, "Supernova 1987A; one year later.",
         LaThuile proceedings, 1988.
 
\bibitem{Kayser} B.Kayser, "On the quantum mechanics of neutrino oscillation",
         Physical Review D, 1981.

\bibitem{Giunti} C.Giunti et al., "When do neutrinos really oscillate?
                                   Quantum mechanics of neutrino oscillations",
         Physical Review D, 1991.

\bibitem{Bilenky} S.Bilenky et al., "Neutrino masses and mixing from neutrino
                                      oscillation experiments",
                   HEP97, 1997.

\bibitem{Mohanty} S.Mohanty, "Production process dependence of neutrino flavor
                              conversion",
         hep-ph/9706328, 1997.

\bibitem{LoSecco} J.LoSecco, "Proceedings of the neutrino mass 
                              mini-conference",
         Telemark Wisconsin, 1980.

\bibitem{Grimus} W.Grimus et al., "Real Oscillations of Virtual Neutrinos",
         UWThPh-1996-17, 1997.

\bibitem{Giannetto} E.Giannetto et al., "Are muon neutrinos faster-than-light
                                         particles?",
         Physics Letters B, 1986.

\bibitem{RecamiWaldyr} E.Recami et al., "Tachyons: May they have a Role in Elementary
                                         Particle Physics?",
         Proceedings of the International school of nuclear physics, 1985.

\bibitem{Eddington} A.S.Eddington, "Space, Time and Gravitation",
         Cambridge, 1920.
                                         


\end{thebibliography}
\end{document}